# The privacy implications of Bluetooth

Vassilis Kostakos
University of Madeira / Carnegie Mellon University
vassilis@cmu.edu

## Abstract

A substantial amount of research, as well as media hype, has surrounded RFID technology and its privacy implications. Currently, researchers and the media focus on the privacy threats posed by RFID, while consumer groups choose to boycott products bearing RFID tags. At the same, however, a very similar technology has quietly become part of our everyday lives: Bluetooth. In this paper we highlight the fact that Bluetooth is a widespread technology that has real privacy implications. Furthermore, we explore the applicability of RFID-based solutions to address these privacy implications.

## Introduction

A substantial amount of research, as well as media hype, has surrounded RFID technology and its privacy implications [1,2]. Currently, researchers and media focus on the privacy threats posed by RFID, while consumer groups choose to boycott products bearing RFID tags. At the same time a very similar technology has quietly become part of our everyday lives: Bluetooth. In a recent study we found that at least 7.5% of observed pedestrians had Bluetooth-enabled mobile devices [3]. Although intended for different uses and purposes, both Bluetooth and RFID allow for mobility and unique identification. It is this characteristic that makes these two technologies similar in terms of the privacy threats they pose.

In this paper we first describe the technical similarities and differences between RFID and Bluetooth. Based on these, we highlight a number of RFID-related privacy threats that equally apply to Bluetooth. We then consider the suitability of RFID-based solutions to these threats, and examine how well they transfer to Bluetooth. We conclude with a real-world case study where Bluetooth was used to identify devices (and presumably their owners) at the scene of a crime.

Our intention is two-fold. First, we wish to highlight the fact that Bluetooth is a widespread technology that has real privacy implications. Second, we want to explore the applicability of RFID-based solutions to address these implications. We conclude that although RFID can *potentially* pose more significant privacy threats, at the moment Bluetooth poses much more considerable threats to privacy.

# RFID vs. Bluetooth

Both RFID and Bluetooth are wireless communication technologies. RFID operates at the 13.56 MHz frequency, and RFID tags have a range of typically up to 30 cm, while self-powered active RFID tags have a range of up to 100m. Bluetooth operates at the 2.45 GHz frequency, with ranges of up to 100 m. While RFID operations typically involve a tag and a reader, Bluetooth communication takes place between Bluetooth-enabled devices. Both technologies, however, rely on unique identifiers for operation.

## *Identification*

Every RFID tag provides a unique identifier (128 bits in newer tags), and additional information up to a few kilobytes may be stored in the tag itself. This identifier is used either as a key (for example to look up the price of a product), or as a mechanism to distinguish between multiple tags. RFID tags can also store information such as the electronic product code assigned to the object, the location or other asset on which the tag is attached, as well as application-related user data.

Bluetooth devices have a 48-bit unique identifier and 24-bit device class descriptor. Additionally, each device can be assigned a "user-friendly" device name (up to 256 characters). Each Bluetooth device provides a number of services, such as file-transfer, serial port, or audio link. These services can be utilised by other nearby Bluetooth-enabled devices. Once a link has been established between two devices, any amount of data can be exchanged.

## *Communication*

Both RFID and Bluetooth operate in a broadcast model for communication. For RFID, a scanner will typically transmit a request signal and wait for a response from all the tags within range. At this point the reader can choose a specific device and query it directly. A very similar process takes place with Bluetooth: an inquiring device hops through a set of prescribed frequencies and transmits a request signal, and any device within range responds to this request. Given the set of responses, the inquiring Bluetooth device can choose to communicate with any specific device within range.

We should highlight an important difference between the two technologies. In RFID there is a clear distinction between tags and readers: tags are "dumb" units with limited functionality whose sole purpose is to be associated with an item in the real world, while readers are smart units whose purpose is to seek RFID tags and query them or modify them. The picture is completely different when we focus on Bluetooth devices. With Bluetooth there is no distinction between tags and readers, as effectively each device can act as both. Furthermore, both parties are equals in communication, as each device can access the other device's services. This fundamental difference between RFID and Bluetooth is a first hint towards the privacy threats posed by Bluetooth.

While end-users are expected to carry RFID tags embedded in everyday objects, there is no intention to equip end-users with RFID scanners, just as users never required barcode readers. Even if users were to be equipped with mobile RFID readers, their scanning potential would be limited due to RFID's limited range. On the other hand, users who own Bluetooth-enabled device effectively own a Bluetooth scanner.

By default, therefore, Bluetooth empowers users to carry out their own scanning of nearby devices. On the other hand, certain knowledge of electronics is required to build an RFID scanner, while off-the-shelf RFID scanners are too narrow in functionality for any end-user to be interested at the moment.

## Privacy threats

The media hype surrounding RFID threats relate to the fact that every single item can potentially be tagged with passive tags – from clothes and shoes to food and medicine. On the other hand, Bluetooth is typically found in communication devices, as it requires a source of power. In a recent study we found that approximately 7.5% of pedestrians had a Bluetooth-enabled device [3]. This study was carried out in the city of Bath, UK, and we expect this figure to be different in other cities and countries.

Realistically, however, Bluetooth is out there on the streets *today*. Most newer phones, PDAs and satellite navigators provide Bluetooth functionality. Therefore, it is worth considering the threats that Bluetooth poses *today*. Here we draw on a summary of RFID privacy threats described in [1]. After describing each RFID threat, we discuss its applicability to Bluetooth.

### *Association threat*

This threat illustrates the fact that a user's identity can be associated with certain items. For example, when buying RFID-tagged goods with a credit card, the seller can link those goods to the credit card, and ultimately, the identity of the buyer.

The exact same threat is posed by Bluetooth technology. Sellers of Bluetooth devices (for example mobile phone operators) can link the Bluetooth unique identifier to the identity of the owner. This can happen at the point of sale. Although at the moment there is no centralised database containing such information, there exists a very similar database linking mobile phone unique identifiers (i.e. IMEI) to individuals. This database could easily be extended to include Bluetooth identifiers.

### *Location threat*

Given the ability to link a tag or item to a person, the person's or item's location can be revealed and tracked. Using RFID this is possible by installing hidden readers at specific locations. This allows for people, or devices, to be identified and tracked.

The exact same threat applies to Bluetooth technology. Any Bluetooth device can be used to identify and track nearby users, and can also be used to determine the location of Bluetooth-enabled artefacts like mobile phones, POS tills, PDAs and satellite navigators. In addition to our own work (e.g. [3]), a number of other projects [4,5] are doing exactly this: tracking users through space using Bluetooth technology. In fact, the first commercial products involving Bluetooth tracking are starting to appear on the market.[1] As we describe in the association threat above, however, although a device can be traced in space and time, at the moment there is no way to link that device to an individual.

---

[1] http://www.blipsystems.com

## Preference threat

In addition to revealing an item's unique identity, RFID tags uniquely identify the manufacturer and product type of the tagged items. This information can be captured and used to create a preference profile of an individual, and even be used to determine the monetary value of these products. For instance, we could detect if a user likes to shop at a specific store, or even identify any specific medicine the user is carrying and thus infer a disease they may have. Similarly, we can calculate the value of the items in a purse.

The exact same threat applies to Bluetooth devices: the provided information can be used to determine the type (or class) of a device (e.g. laptop, desktop, mobile phone) and the manufacturer of the product (e.g. Nokia, Sony-Ericsson, Apple). Additionally, advanced fingerprinting techniques can determine the exact model of a mobile phone.[2] Such techniques also enable us to derive estimates of the monetary value of the scanned products.

## Constellation threat

Users carrying dozens of RFID-tagged items can be seen to have a digital "shadow" or "constellation." A constellation can be used to track individuals without necessarily knowing their identity.

This is also true of Bluetooth, but to a smaller extent. Users typically carry a handful of Bluetooth-enabled devices, such as a mobile phone, PDA, headset, or satellite navigator. These, however, still create a digital shadow that can be tracked and monitored. Another version of this threat is to derive associations between people: a Bluetooth constellation might describe a group of people, and the behaviour of this constellation can allow us to infer properties of this group.

## Transaction threat

Transactions between individuals can be tracked by observing changes in constellations. When an RFID-tagged item moves from one constellation to another, we can determine that a transaction between those individuals took place.

The exact same threat holds for items forming Bluetooth constellations, although to a smaller extent. The limited scope for transaction threats in Bluetooth is due to the small number of devices that an individual may carry.

## Breadcrumb threat

As a consequence of the association threat, a user can e constantly linked to RFID-tagged items they purchased. This association does not necessarily break when the user discards those items. The breadcrumb threat arises when discarded items are used, for example, to commit a crime, in which case the original owner might be implicated.

In terms of Bluetooth this is a perfectly plausible scenario, however it is limited by the fact that Bluetooth items are not necessarily discarded. It is still possible, however, that a user discards, for example, a mobile phone, which in turn is used by a criminal at the scene of a crime.

---

[2] http://trifinite.org/trifinite_stuff_blueprinting.html

# Solutions

The above threats, although initially conceived in relation to RFID technology, also apply to Bluetooth technology. Actually, in most cases the threats are exaggerated when considering Bluetooth technology because of Bluetooth's longer range and the ease and availability of Bluetooth scanning. We now consider some of the solutions proposed in relation to RFID [2], and examine their applicability to Bluetooth.

## *Discarding or destruction*

RFID tags can be completely discarded after they have served their purpose. For example, it has been proposed that RFID tags are destroyed at the point of sale. Similarly, users can remove a tag from the products they buy, provided that these tags are ease to locate and remove.

This solution, however, does not transfer to Bluetooth at all. The reason is that whereas RFID is mostly useful *up to* the point of sale, Bluetooth is really useful *after* the point of sale. Destroying the Bluetooth functionality of a mobile device would actually damage the device and limit its capabilities. Most mobile devices rely on Bluetooth for establishing wireless links, exchanging files, and synchronising. Furthermore, while RFID tags may be visible and easily removed, Bluetooth is typically embedded in products, and thus cannot easily be destroyed without damaging the product itself.

## *Deactivation or suppressing*

Elaborate security mechanisms have been proposed to temporarily deactivate or suppress RFID tags [2]. The complexity of this process in RFID is due to the fact that a tag cannot be deactivated directly, but instead a reader must send it a deactivation command. This, in turn, raises concerns of unauthorised readers being able to deactivate tags. Therefore, security measures involving passwords are used to ensure that only authorised readers can deactivate a tag.

Conceptually, Bluetooth also uses deactivation as a privacy measure. In practice, however, the process is slightly different. Unlike RFID, users can directly interact with the Bluetooth functionality of their mobile device. Users can either

- completely switch off their device,
- selectively disable Bluetooth on their device, or
- configure their Bluetooth to operate in "stealth" mode, which effectively suppresses the unique identifier.

The first two options do not allow for any Bluetooth communication to take place, while the third option allows for Bluetooth communication to take place with specific and known devices, also referred to as "paired" devices.

There are two caveats, however, to these privacy measures. First, our studies have shown that a considerable portion of the public (7.5%) does not deactivate or suppress their Bluetooth devices [3]. This is partly due to the culture, at least in the UK, of giving one's device funny names, such that others can see them when they carry out Bluetooth scans. Furthermore, it could simply be due to the lack of knowledge and understanding of how Bluetooth operates. Secondly, once an individual's identity has been linked to a Bluetooth device (association threat), it becomes trivial to scan for that specific individual *even if* their device operates in stealth mode. This is possible because stealth mode assumes that other parties do not know a device's unique identifier. Given that unique identifier, however, stealth mode offers no privacy protection.

## Renaming

Another approach to dealing with RFID privacy concerns is to periodically change the unique identifiers that a tag emits. Thus, only readers with knowledge of the renaming sequence can keep track of the specific tags. A way to achieve this is for tag and reader to, for example, share a predefined set of identifiers through which a tag cycles over its lifetime.

This solution can potentially work with Bluetooth, although it becomes much more complicated since, with Bluetooth, renaming must take place on both communicating parties. For this solution to work with Bluetooth, each device must keep track of the renaming cycle on every other device it talks to, and visa versa. This is further complicated by the fact that Bluetooth devices can operate in networks of up to 8 devices, also known as piconets. An elaborate mechanism is required to allow all communicating parties to be updated of the name changes that take place within the piconet.

## Exploiting physical properties

Many of the RFID-related privacy issues can be addressed by making sure that the scanner attempting to read a tag is within a very close range. Doing so avoids the issues caused by hidden or covert scanners. This can be achieved by installing certain circuits within the tags that effectively limit the actual range of RFID.

This approach is by definition not applicable to Bluetooth. The range of Bluetooth is an extremely useful feature, and minimising its range would render Bluetooth useless. An approach that might be more practical would involve the installation of a knob, similar to the volume knobs found in walkmans, which would dynamically adjust the strength of the Bluetooth signal.[3] This would allow users to extend their signal as much as is required. However, anyone within that distance would still be able to instantiate all the threats we have presented.

## Non-RFID solutions

The solutions we have presented so far originate in research on RFID privacy threats. Some of these solutions transfer to Bluetooth (i.e. deactivation, renaming, exploiting physical properties), while other solutions cannot be transferred to Bluetooth (i.e. discarding and destruction). Here we present some further solutions that are only applicable to Bluetooth communications and we have developed with Bluetooth in mind.

Our first suggested solution is the *hit counter*. At the moment, the Bluetooth protocol allows a device to be notified when another device has inquired the list of services, but not when other devices "discovered" it. However, a slight modification of the Bluetooth stack would allow each device to keep track of how many times it has been discovered, much like the hit counter on web pages. This number can be displayed to users, who can thus be made more aware of Bluetooth activity. Users can thus choose to disable Bluetooth on their devices when there appears to be unknown and/or systematic Bluetooth activity.

---

[3] http://www.blipsystems.com/Default.aspx?ID=684

Another suggested solution is the *guest book*. Building on the hit counter solution, a device can keep a record of the identifiers of the devices that have discovered it. The user would then be able to inspect the list of those devices, and identify traces of unauthorised activity.

Our final solution to protecting privacy is an adaptation of the renaming solution. We propose that Bluetooth can rely on the "user-friendly" name, as opposed to the unique identifiers. This solution is quite similar to the idea of using self-assigned IP addresses for peer-to-peer WiFi communication. Typically, Bluetooth operations take place between two devices, thus it can be relatively straightforward for the user or users to make sure that each device is given a unique name. Additionally, each device could self-assign a name (or part of the name) periodically, or for each Bluetooth activity. The drawback of this solution is that pairings between devices can only last until one of the two devices changes its name. This also means frequent tasks (such as synchronisation of data using Bluetooth) would require an extra step of device discovery.

## Case study: Bluetooth CSI

We illustrate the reality of Bluetooth privacy threats by describing a case study where we used Bluetooth to carry out our own, informal, crime scene investigation (CSI). In the early morning of November 29, 2006, the police in Bath, UK, were informed of a teenager dying after having fallen from the top of a building in the city centre. Early reports mentioned a second individual who subsequently fled the scene of the incident. The following day, the police appealed to the general public for anyone with any information to come forward.

This incident took place at the building where one of our Bluetooth scanners is located. This scanner, which is part of a network of scanners across the city of Bath, is used to record mobility traces of people, much like the work carried out by similar projects [4,5]. The night of the incident, our scanner was fully operational, and constantly recorded the presence of any Bluetooth device within its range.

Shortly after we began our analysis of the data from the night of the incident, the police ended the enquiry. The unlucky teenager and his friend were drunk, they decided to climb to the top of the building and in an unfortunate moment the teenager lost his balance and fell, landing on the sidewalk.

In our limited analysis of the data we were able to identify a number of devices present at the scene of the incident shortly before and shortly after the teenager fell. Because there is no way to link a Bluetooth identifier to someone's identity, our data cannot be used to directly implicate anyone. Additionally, we can only be certain about the device being present at the location, not the person. However, our data could be used to narrow down the search for potential eyewitnesses. For example, the police could have achieved this by sweeping the streets and identifying people whose device identifier matched our records.[4]

---

[4] Although our data is stored using a one-way hash function, the same function can be applied in real time by the police to generate a match.

In addition to identifying a set of devices that appeared at the scene of the incident, we were able to cross-reference our database and identify the appearance patterns of those people. This subsequently gave us further information about each device. For instance, one of the devices must belong to a taxi driver or delivery van, as it is a satellite navigation unit and in our database appears many times throughout the day or night at this location. We should mention that there is a taxi rank nearby this location. A second device must belong to an individual such as a street sweeper or security guard, since in our database this device appears in a number of locations across the city centre, and always between 10pm and 3am.

We have not been able to verify the findings of our data analysis. Despite this, we feel that very useful information can be extracted from our dataset, as well as any proximity-based traces database. Patterns of behaviour can help identify individuals, their occupation, as well as predict where they are going to be.

## Conclusion

While RFID has captured the attention of both the research and consumer communities, Bluetooth has quietly become one of the most widely used technologies found on the streets today. In this article we describe the remarkable similarities in the threats posed by RFID and Bluetooth. Furthermore, while a number of mechanisms have been developed to curb those threats in relation to RFID, few of these mechanisms transfer well to Bluetooth. In fact, Bluetooth's increased range and ease of scanning greatly increases those threats. We demonstrate the reality of the privacy threats posed by Bluetooth in a case study involving a lethal accident. In table 1 we summarise the RFID threats we have discussed in this paper, and their applicability to Bluetooth. Similarly, in Table 2 we summarise the proposed solutions and their applicability to Bluetooth.

| Threat | RFID | Bluetooth |
|---|---|---|
| Association threat | ✓ | ✓ (requires database) |
| Location threat | ✓ | ✓ |
| Preference threat | ✓ | ✓ |
| Constellation threat | ✓ | ✓ (applies to groups) |
| Transaction threat | ✓ | ✓ (limited) |
| Breadcrumb threat | ✓ | ✓ (limited) |

Table 1. Privacy threats using RFID and Bluetooth.

| Solution | RFID | Bluetooth |
|---|---|---|
| Discard | ✓ | |
| Destroy | ✓ | |
| Deactivate/suppress | ✓ | ✓ |
| Rename | ✓ | ✓ (considerable adaptation) |
| Exploit physical properties | ✓ | ✓ |
| Hit counter | | ✓ (protocol patch) |
| Guest book | | ✓ (protocol patch) |
| User-friendly name only | | ✓ (protocol adaptation) |

Table 2. Privacy preserving strategies and their applicability to RFID and Bluetooth.

Although the Bluetooth threats we have described are real, and the penetration of Bluetooth has reached a considerable portion of the public, very few cases have been documented where Bluetooth was used to instantiate any of the threats we have described here. Perhaps one of the stories that gained some publicity was Paris Hilton's famous hacking of her phone using Bluetooth.[5] Ironically, however, this attack exploited weaknesses in the Bluetooth security model, and had nothing to do with the threats posed by the unique identifiers described here.

Instead of causing havoc, Bluetooth's "leakiness" has resulted in a culture of playful exploration and exchange of media, at least in the UK. Typically bored teenagers, as well as adults, engage in various games using Bluetooth, ranging from the simple exchange of messages in night clubs to playing Bluetooth "hide and seek" on ships: taking a photograph of a location on the ship and transmitting it to the other person using Bluetooth. Despite all the games, however, the threats persist.

To conclude, we should mention that many popular communication technologies, such as WiFi and GPRS, also involve the use of unique identifiers. Unlike RFID's truly pervasive potential, Bluetooth and other similar technologies are not intended to be used on almost every conceivable object. Despite this, technologies such as Bluetooth are quite popular *already*; it is worth considering, therefore, the privacy threats they pose, and the mechanisms to address those threats.

---

[5] http://www.cbsnews.com/stories/2005/02/21/scitech/pcanswer/main675320.shtml